\documentclass[12pt,preprint]{aastex}
\def\degpoint{\ifmmode ^{\rm{o}}\!. \else $^{\rm{o}}\!.$\fi}
\newcommand{\degrees}{$^{\rm{o}}$}
\newcommand{\ms}{\mbox{m\,s$^{-1}$}}
\newcommand{\kms}{\mbox{km \ s$^{-1}$}}
\newcommand{\Msun}{\mbox{M$_{\odot}$}}
\newcommand{\Rsun}{\mbox{R$_{\odot}$}}
\newcommand{\Mjup}{\mbox{M$_{\rm Jup}$}}

\newcommand{\gtsimeq}{\raisebox{-0.6ex}{$\,\stackrel
         {\raisebox{-.2ex}{$\textstyle >$}}{\sim}\,$}}

\begin{document}

\title{The Anglo-Australian Planet Search. XXIII. Two New Jupiter 
Analogs }

\author{Robert A.~Wittenmyer\altaffilmark{1,2}, Jonathan 
Horner\altaffilmark{1,2}, C.G.~Tinney\altaffilmark{1,2}, 
R.P.~Butler\altaffilmark{3}, H.R.A.~Jones\altaffilmark{4}, Mikko 
Tuomi\altaffilmark{4,5},G.S.~Salter\altaffilmark{1,2}, 
B.D.~Carter\altaffilmark{6}, F.~Elliott Koch\altaffilmark{7}, 
S.J.~O'Toole\altaffilmark{8}, J.~Bailey\altaffilmark{1,2}, 
D.~Wright\altaffilmark{1,2} }

\altaffiltext{1}{School of Physics, University of New South Wales, 
Sydney 2052, Australia}
\altaffiltext{2}{Australian Centre for Astrobiology, University of New 
South Wales, Sydney 2052, Australia}
\altaffiltext{3}{Department of Terrestrial Magnetism, Carnegie
Institution of Washington, 5241 Broad Branch Road, NW, Washington, DC
20015-1305, USA}
\altaffiltext{4}{University of Hertfordshire, Centre for Astrophysics
Research, Science and Technology Research Institute, College Lane, AL10
9AB, Hatfield, UK }
\altaffiltext{5}{University of Turku, Tuorla Observatory, Department of 
Physics and Astronomy, V\"ais\"al\"antie 20, FI-21500, Piikki\"o, 
Finland}
\altaffiltext{6}{Computational Engineering and Science Research Centre, 
University of Southern Queensland, Toowoomba, Queensland 4350, 
Australia}
\altaffiltext{7}{San Diego State University, Physics Department, 5500 
Campanile Drive San Diego, CA 92182-1233, USA}
\altaffiltext{8}{Australian Astronomical Observatory, PO Box 915,
North Ryde, NSW 1670, Australia}

\email{
rob@phys.unsw.edu.au}

\shortauthors{Wittenmyer et al.}

\begin{abstract}

\noindent We report the discovery of two long-period giant planets from 
the Anglo-Australian Planet Search.  HD\,154857c is in a multiple-planet 
system, while HD\,114613b appears to be solitary.  HD\,114613b has an 
orbital period $P=10.5$ years, and a minimum mass m~sin~$i$ of 
0.48\Mjup; HD\,154857c has $P=9.5$ years and m~sin~$i$=2.6\Mjup.  These 
new data confirm the planetary nature of the previously unconstrained 
long-period object in the HD\,154857 system.  We have performed detailed 
dynamical stability simulations which show that the HD\,154857 
two-planet system is stable on timescales of at least $10^8$ yr.  These 
results highlight the continued importance of ``legacy'' surveys with 
long observational baselines; these ongoing campaigns are critical for 
determining the population of Jupiter analogs, and hence of those 
planetary systems with architectures most like our own Solar system.

\end{abstract} 

\keywords{planets and satellites: individual (HD\,114613, HD\,154857) -- 
planets and satellites: detection -- techniques: radial velocities -- 
planets and satellites: dynamical evolution and stability }

\section{Introduction}

A major theme that has unified exoplanet searches for more than 20 years 
is the question of how common (or rare) our own Solar system is. The 
\textit{Kepler} spacecraft, which continuously monitored over 100,000 
stars for tiny eclipses caused by orbiting planets \citep{borucki10}, 
has provided exquisite data which have revolutionised our understanding 
of the frequency of Earth-size planets in short-period orbits 
\citep{howard12, fressin13}.  However, \textit{Kepler} alone cannot give 
us a complete picture of the occurrence rate of planetary systems like 
our own, with rocky inner planets and one or more gas giant planets 
(``Jupiter analogs'') at orbital distances $a\gtsimeq$3\,AU.  There is 
more to a system being Solar System-like than having a single planet in 
a potentially habitable orbit.  The detection of a Jupiter analog is a 
second key component in determining whether an exoplanetary system is 
Solar system-like.  Over the years, many arguments have been put forth 
to suggest that such external giant planets might be a necessity for a 
potentially habitable exo-Earth to be a promising location for the 
development of life \citep{hj10}.  Although the role of such planets 
acting as a shield from an otherwise damaging impact regime has come 
into question \citep[e.g.][]{hj08, hjc10, lewis13}, a number of other 
potential benefits are thought to accrue from the presence of 
Jupiter-analogs.  For example, Jupiter-like planets have been proposed 
as a solution to the question of the origin of Earth's water. Current 
models of planetary formation suggest that the Earth formed in a region 
of the proto-planetary disk that was far too warm for water to condense 
from the gas phase.  As such, it is challenging to explain the origin of 
our planet's water without invoking an exogenic cause.  The formation 
and evolution of the giant planets, beyond the snow line, offers a 
natural explanation for the delivery of volatiles from the cold depths 
of a planetary system to planets that move on potentially habitable 
orbits (e.g.~Horner et al. 2009; Horner \& Jones 2010b, and references 
therein).  The detection of a Jupiter-analog is therefore both a second 
key component in determining whether an exoplanetary system is Solar 
system-like, and a potential marker that the planets in that system 
might be promising targets for the future search for life beyond the 
Solar system.

The Anglo-Australian Planet Search (AAPS) has been in operation for 15 
years, and has achieved a long-term radial-velocity precision of 3 \ms\ 
or better since its inception, which is enabling the detection of 
long-period giant planets.  To date, the AAPS has discovered six 
Jupiter analogs: HD\,70642b \citep{carter03}, HD\,160691c \citep{m04}, 
HD\,30177b \citep{butler06a}, GJ\,832b \citep{bailey09}, HD\,134987c 
\citep{jones10}, HD\,142c \citep{142paper}.  Here, we have defined a 
Jupiter analog as a giant planet which has ended up near its formation 
location, beyond the ice line, with $a>3$\,AU.  Recently, the AAPS has 
shifted its priority to the detection of these Jupiter analogs.  The 
observing strategy and target list have been modified, with the aim of 
producing an accurate and precise determination of the frequency of 
Jupiter-like planets in Jupiter-like orbits \citep{jupiters,witt13}.  
The modified target list includes stars with long-term velocity 
stability such that Jupiter analogs can be robustly excluded 
\citep[e.g.][]{limitspaper, etaearth}, as well as those stars with 
as-yet-incomplete orbits suggestive of long-period giant planets.  In 
this paper, we report the discovery of two such Jupiter analogs with 
complete orbits.  HD\,154857 is already known to host a 1.8\Mjup\ planet 
with an orbital period of about 400 days \citep{m04}; a residual 
velocity trend indicated a much longer-period object, as noted in the 
discovery work and in \citet{otoole07}.

This paper is organized as follows: Section 2 briefly describes the 
observational details and stellar parameters, and Section 3 details the 
orbit fitting process and gives the parameters of the two new planets.  
In Section 4, we present a dynamical stability analysis of the 
HD\,154857 two-planet system, and we give our conclusions in Section 5.

\section{Observations and Stellar Parameters}

AAPS Doppler measurements are made with the UCLES echelle spectrograph 
\citep{diego:90}.  An iodine absorption cell provides wavelength 
calibration from 5000 to 6200\,\AA.  The spectrograph point-spread 
function (PSF) and wavelength calibration are derived from the iodine 
absorption lines embedded on the spectrum by the cell 
\citep{val:95,BuMaWi96}.  The result is a precise Doppler velocity 
estimate for each epoch, along with an internal uncertainty estimate, 
which includes the effects of photon-counting uncertainties, residual 
errors in the spectrograph PSF model, and variation in the underlying 
spectrum between the iodine-free template and epoch spectra observed 
through the iodine cell.  All velocities are measured relative to the 
zero-point defined by the template observation.  For HD\,114613, a total 
of 223 AAT observations have been obtained since 1998 Jan 16 
(Table~\ref{114613vels}) and used in the following analysis, 
representing a data span of 5636 days (15.4 yr).  The mean internal 
velocity uncertainty for these data is 0.94\,\ms.  HD\,154857 has been 
observed 42 times since 2002 April (Table~\ref{154857vels}), for a total 
time span of 4109 days (11.3 yr) and a mean internal uncertainty of 
1.71\,\ms.

HD\,114613 (HR\,4979; HIP\,64408) is an inactive G-type star, listed as 
a dwarf by \citet{torres06}, though its surface gravity is more 
indicative of a slightly evolved subgiant (Table~\ref{114star}).  It is 
a nearby and bright star ($V=4.85$) with a somewhat super-solar 
metallicity $[Fe/H]\sim$0.19.  HD\,154857 has been classified as a G5 
dwarf \citep{houk75}.  However, all recent measurements of its surface 
gravity show that this star is a subgiant (Table~\ref{154star}).  There 
is some confusion as to the mass: \citet{vf05} give two disparate mass 
estimates, 2.10$\pm$0.31\Msun\ derived from spectroscopic analysis, and 
1.27$^{+0.35}_{-0.29}$\Msun\ from interpolation on a grid of Yonsei-Yale 
isochrones.  For most stars in their sample, the two mass estimates 
agreed within $\sim$10\%, but for HD\,154857, they differ by almost a 
factor of 2.  The more recent analysis by \citet{takeda07} yields an 
intermediate value of 1.718$^{+0.03}_{-0.022}$\Msun, which we adopt in 
this paper.

\section{Orbit Fitting and Planetary Parameters}

\subsection{HD 114613}

\begin{figure}
\plotone{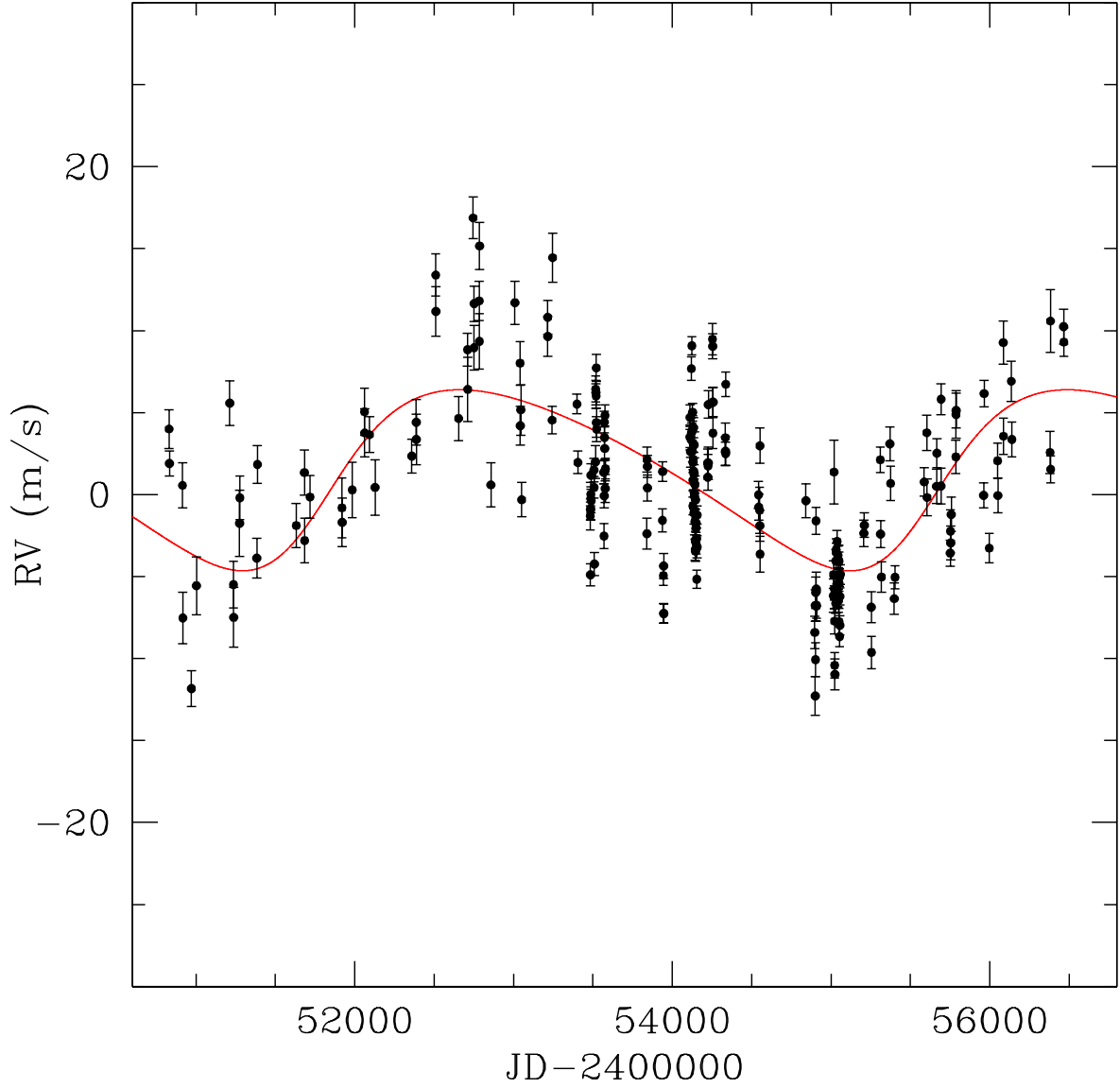}
\caption{Keplerian orbit fit for HD\,114613b.  The planet has completed 
about 1.5 cycles, and the AAT data show a residual rms scatter of 
3.9\,\ms.}
\label{114fit}
\end{figure}

\begin{figure}
\plottwo{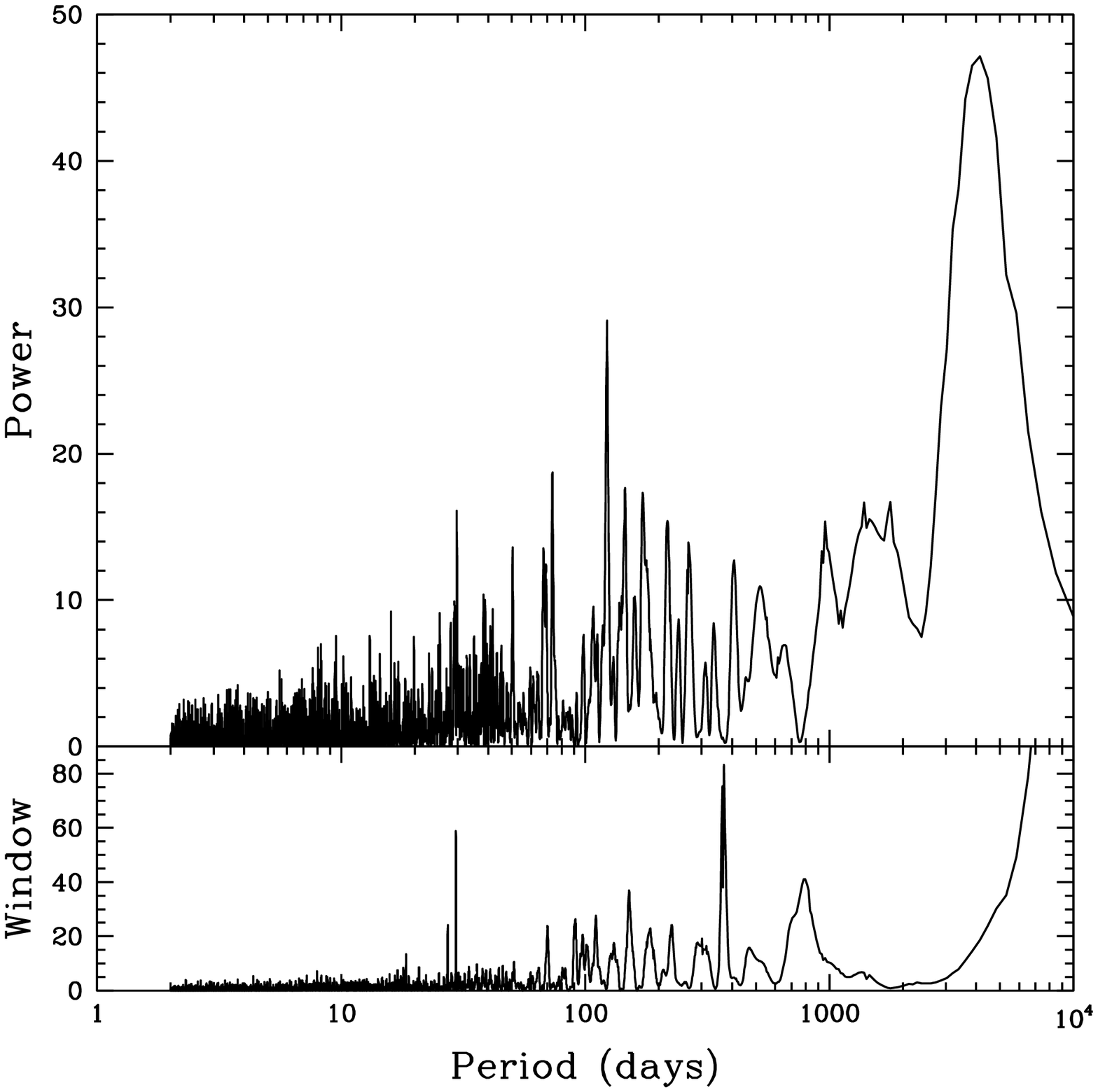}{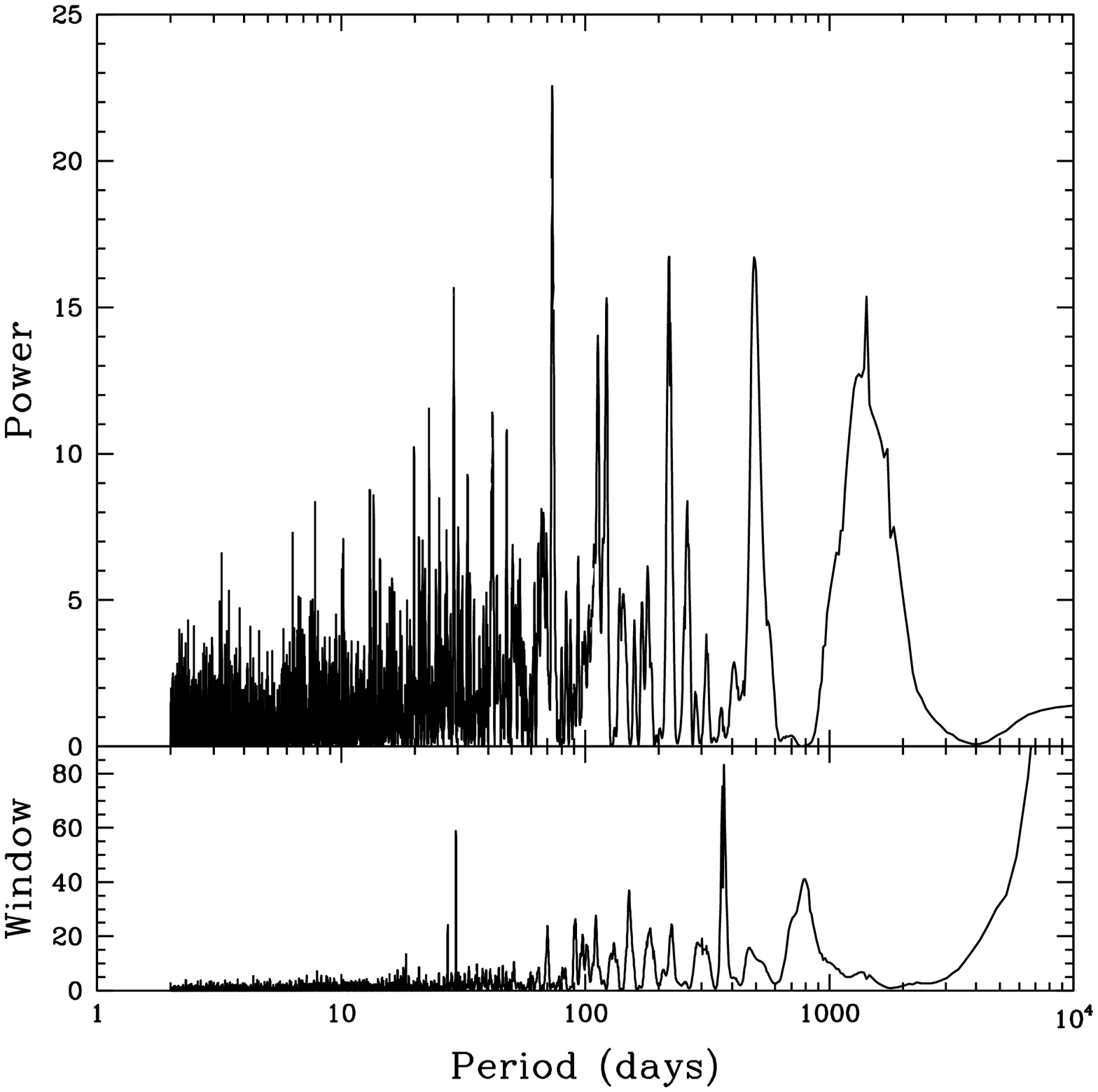}
\caption{Left: Generalized Lomb-Scargle periodogram of 223 AAT 
observations of HD\,114613. A highly significant peak is present near 
4000 days.  Right: Periodogram of the residuals after fitting and 
removing the long-period planet with the parameters given in 
Table~\ref{planetparams}; several peaks remain which may indicate 
further planets. }
\label{pgram1}
\end{figure}

HD\,114613 has been observed by the AAPS for the full 15 years of its 
operation.  A long-period trend had been evident for several years, and 
in 2011, the trend resolved into a complete orbital cycle.  We have 
since continued to observe HD\,114613 to verify that the $\sim$11 yr 
orbit was indeed turning around. Figure~\ref{pgram1} shows the 
Generalized Lomb-Scargle periodogram \citep{zk09} of the 223 AAT 
observations.  This type of periodogram weights the input data by their 
uncertainties, whereas the traditional Lomb-Scargle method 
\citep{lomb76,scargle82} assumes uniform, Gaussian distributed 
uncertainties.  To assess the significance of any signals appearing in 
these periodograms, we performed a bootstrap randomization process 
\citep{kurster97}.  This randomly shuffles the velocity observations 
while keeping the times of observation fixed.  The periodogram of this 
shuffled data set is then computed and its highest peak recorded.  The 
longest-period peak near 4000 days is well-defined and highly 
significant, with a bootstrap false-alarm probability less than 
$10^{-5}$.  The next-highest peaks are at 122 and 1400 days, 
respectively.  We fit these data with a single, long-period Keplerian 
using the \textit{GaussFit} \citep{jefferys87} nonlinear least-squares 
minimization routine.  Jitter of 3.42 \ms\ \citep{wright05, monster} was 
added in quadrature to the uncertainties at each epoch prior to orbit 
fitting.  A single-planet fit yields a period $P=3825\pm$106 d, 
$K=5.4\pm$0.4 \ms, and $e=0.25\pm$0.08 (Table~\ref{planetparams}), 
making this planet a Jupiter analog \citep{jupiters}, with a minimum 
mass m sin $i$ of 0.5\Mjup, and an orbital period of 10.7 years 
(Figure~\ref{114fit}).  The rms about the one-planet fit is 3.9\,\ms, 
and the periodogram of the residuals to this fit is shown in the right 
panel of Figure~\ref{pgram1}, showing a number of peaks ranging from 28 
to $\sim$1500 d.



\begin{figure}
\plottwo{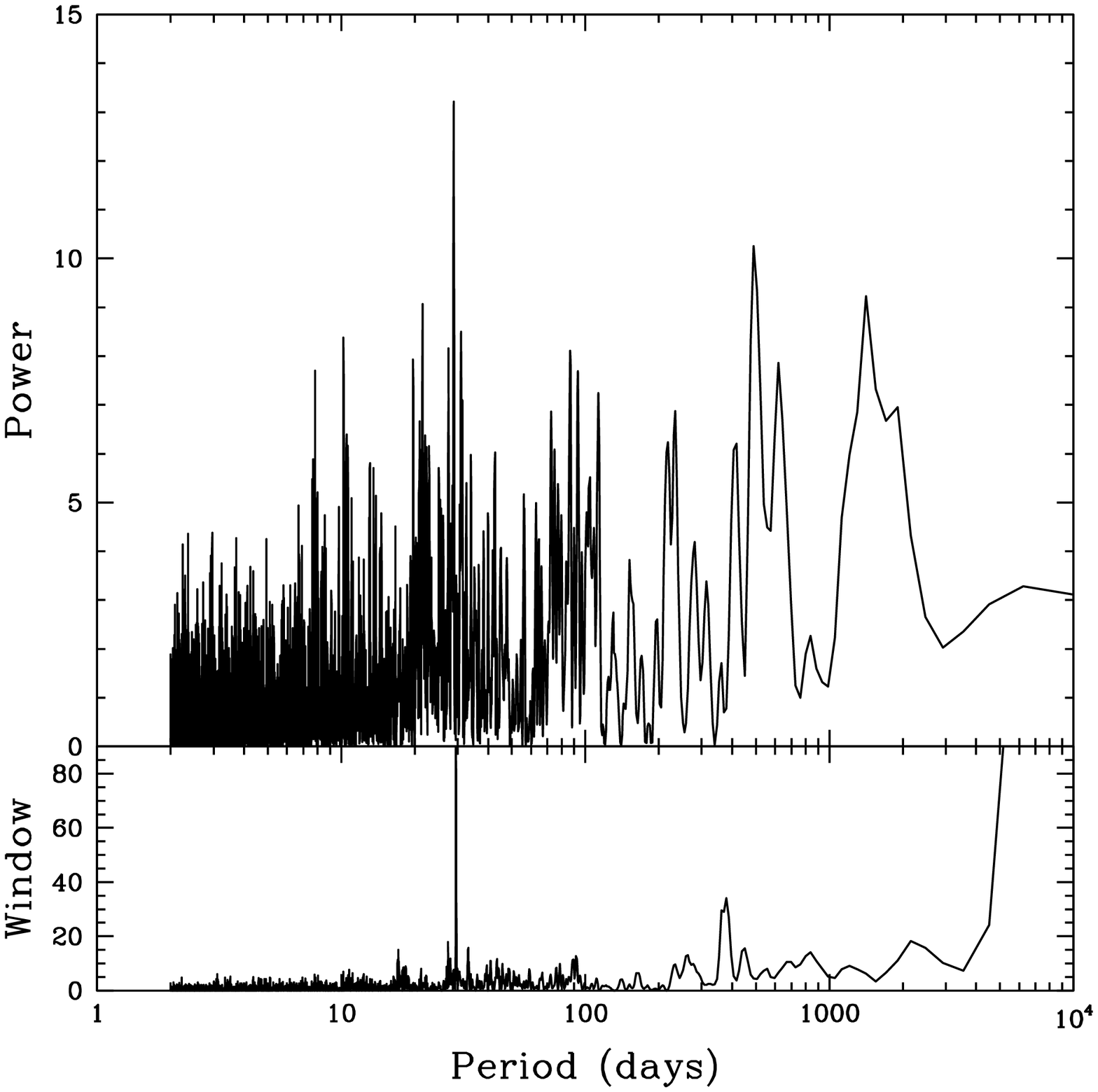}{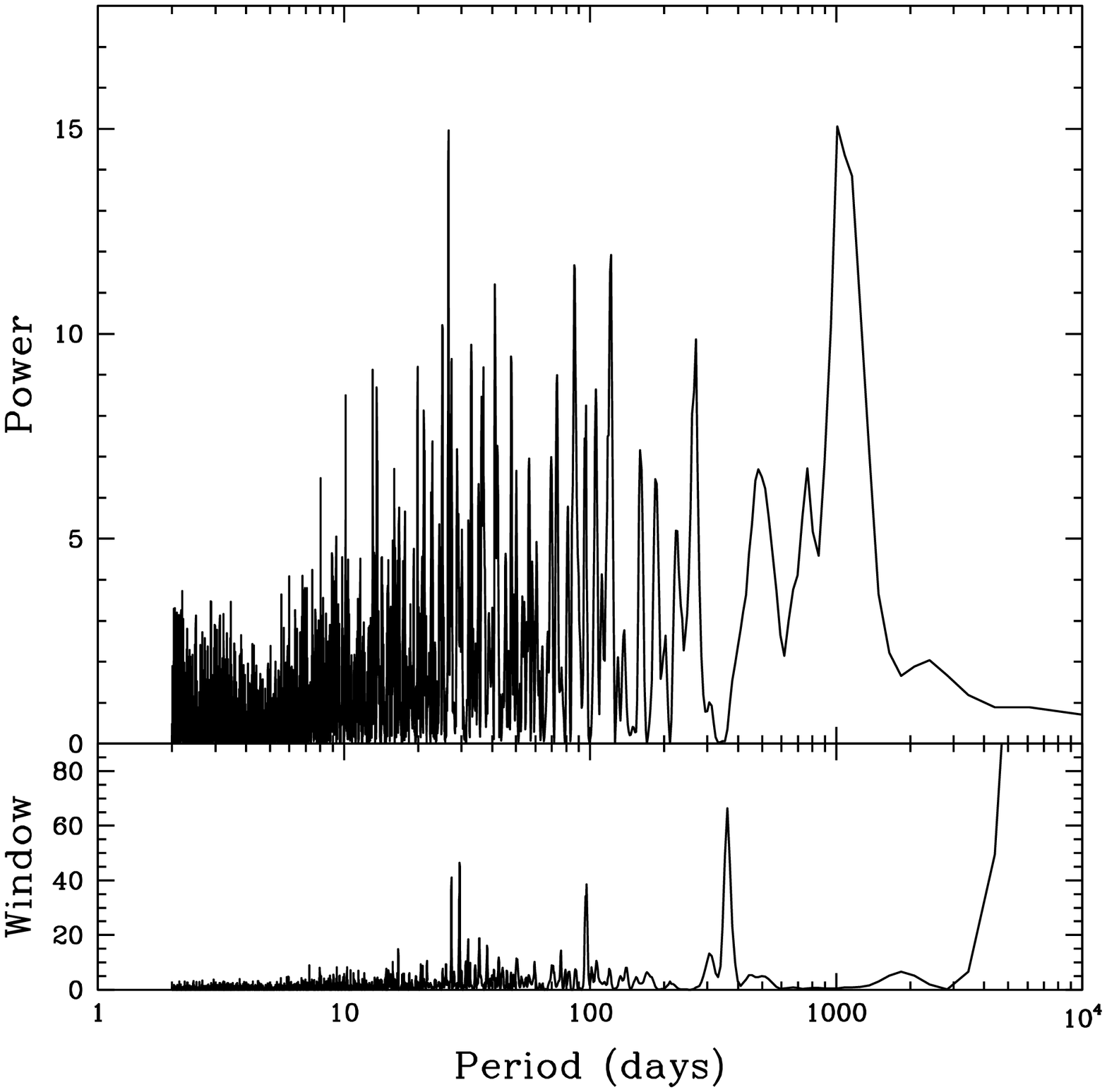}
\caption{Periodograms of the residuals to a single-planet fit for 
HD\,114613.  Left: First eight years ($N=77$). Right: Second eight years 
($N=80$), excluding the high-cadence observing runs in 2007 and 2009. }
\label{split}

\end{figure}

There is structure evident in this residual periodogram 
(Figure~\ref{pgram1}, right panel), so we examined the residuals for 
additional Keplerian signals.  One way of determining the veracity of 
such signals is to examine the data by seasons or subsets.  This can 
disentangle true planetary signals (which would consistently appear in 
all subsets) from stochastic signals such as stellar rotational 
modulation \citep{alphacen, hatzes13}.  We divided the residuals to the 
one-planet fit into two eight-season chunks.  HD\,114613 was observed 
intensely in 2007 and 2009 as part of the AAT ``Rocky Planet Search'' 
campaigns \citep{monster, 16417paper}, in which 24-30 bright stars were 
observed nightly for 48 continuous nights in search of short-period 
planets.  It is possible that such a density of observational data may 
skew the false-alarm probabilities when evaluating potential additional 
signals.  We thus removed the 66 epochs from the two ``Rocky Planet 
Search'' campaigns -- this resulted in the 8-year halves containing 77 
and 80 observations, respectively\footnote{Fitting a single planet with 
this shortened dataset gives parameters within 1$\sigma$ of those given 
in Table~\ref{planetparams}, showing that the exclusion of those data do 
not affect our conclusions about the long-period planet.}.  Periodograms 
of the two halves are shown in Figure~\ref{split}; visual inspection 
reveals that they are markedly different.

Table~\ref{boot} shows the false-alarm probabilities obtained from 
10,000 such realizations on each half of the 1-planet residuals.  No 
periodicity is consistently significant in both subsets, with the 
possible exception of that near 27-29 days -- however, this is 
worryingly close to both the 33-day rotation period of the star 
\citep{saar97} and the lunar month (at which the sampling of 
radial-velocity observations is well-known to impart spurious 
periodicities e.g. Dawson \& Fabrycky 2010, Wittenmyer et al. 2013b).  
While it is tempting to consider a second planet near 1400 days, as 
found in the raw-data periodogram (Figure~\ref{pgram1}), we see that 
this signal is simply not evident in the first 8 years of observations.  
As suggested by \citep{hatzes13} for the proposed planet orbiting Alpha 
Centauri B, we rephrase his sentiments to express that any shorter 
period periodic signal which is evident in one subset of our data should 
also be evident in other subsets or seasons of the data.  As both 
HD\,114613 8-year subsets have ample time coverage and data quantity 
($N=77$) to sample the candidate periods listed in Table~\ref{boot}, we 
can use these results to conclude that there is not yet sufficient 
evidence for additional planetary signals in our data for HD\,114613.

\subsection{HD 154857}

\begin{figure}
\plotone{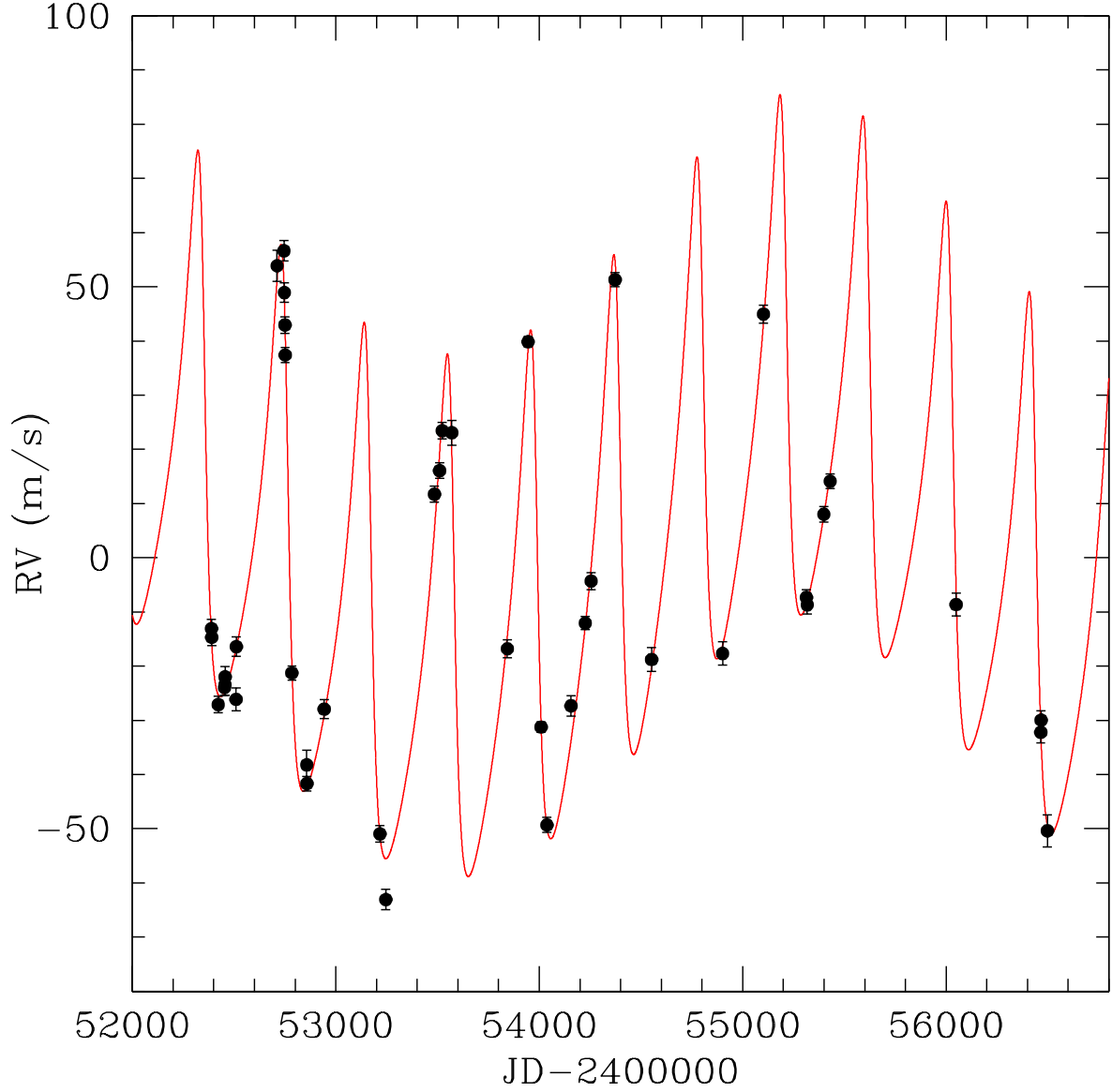}
\caption{Two-planet Keplerian model and AAT data for the HD\,154857 
system. The rms about this fit is 3.2\ms. }
\label{154fit}

\end{figure}

\begin{figure}
\plottwo{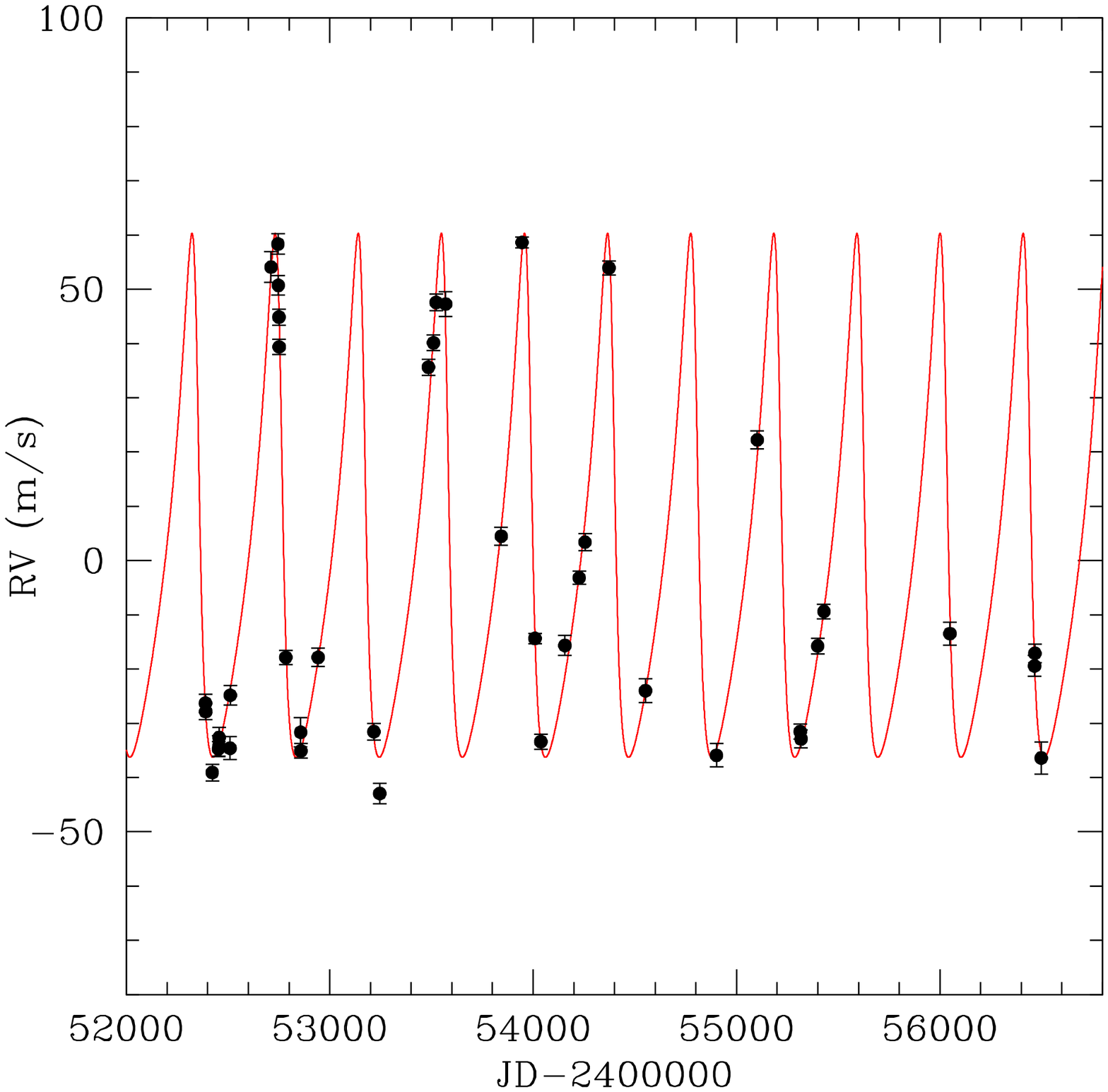}{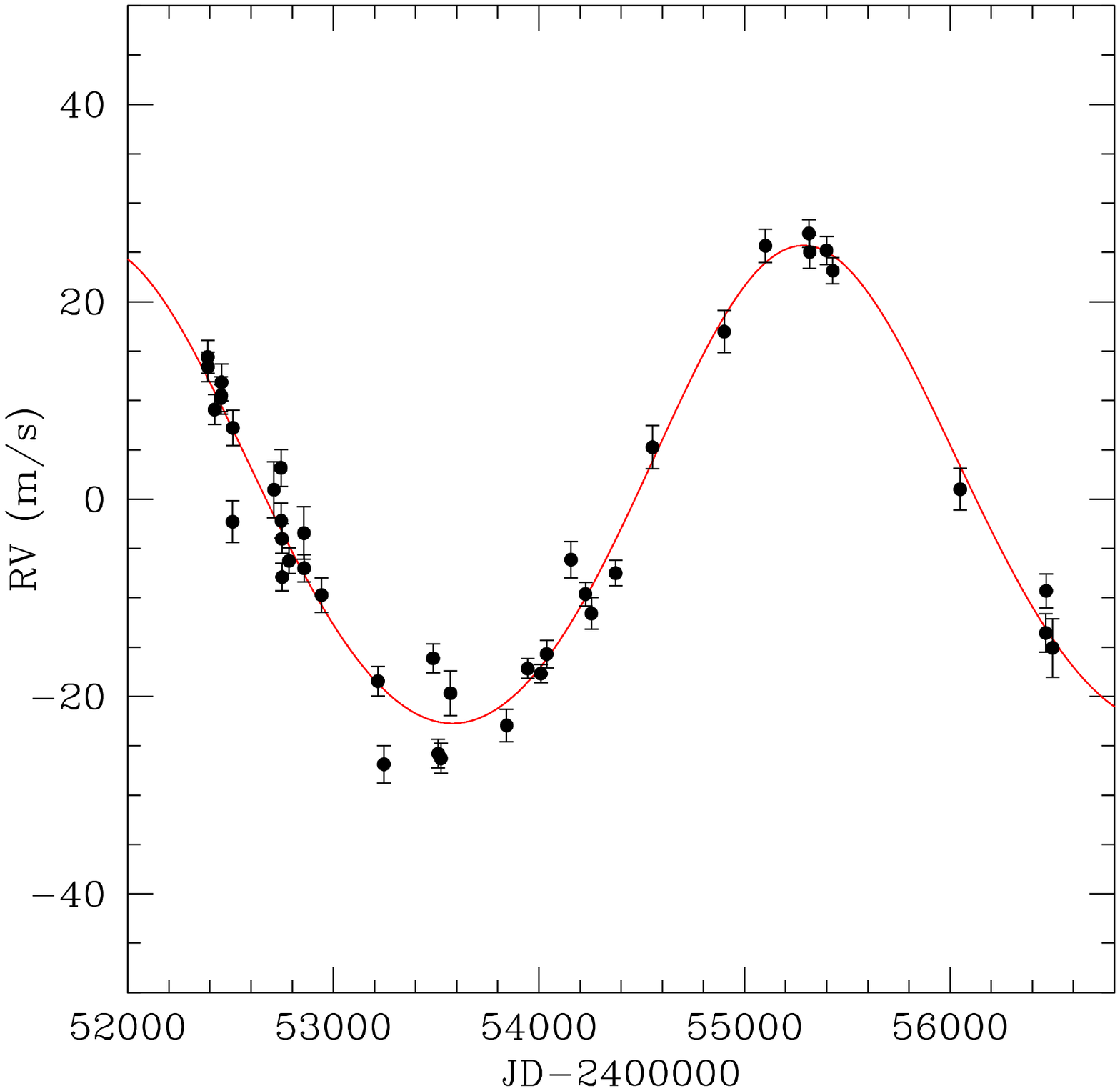}
\caption{Left: Data and model fit for HD\,154857b; the signal of the 
outer planet has been removed. Right: Same, but for HD\,154857c after 
removing the inner planet. }
\label{154eachone}

\end{figure}

The presence of a planet orbiting HD\,154857 was first reported by 
\citet{m04}, who noted that the AAPS data were best fit with the 
$\sim$400-day planet and a linear trend, indicating a more distant body.  
Additional data presented in \citet{otoole07} refined the planet's 
parameters and attempted to constrain the outer object's orbit since the 
residual velocity trend had begun to show curvature.  They determined a 
minimum orbit with period 1900 days and $K\sim$23 \ms.  Now, with a 
further 6 years of AAT data, the outer planet has completed an orbit and 
a double-Keplerian model converges easily.  We used \textit{GaussFit} as 
described above to fit the two planets, first adding jitter of 2.6\,\ms\ 
in quadrature to the uncertainties (after O'Toole et al. 2007).  The 
best-fit parameters are given in Table~\ref{planetparams}; the outer 
planet is a Jupiter analog moving on an essentially circular orbit with 
$P=9.5$\,yr ($a=5.36$AU) and m~sin~$i=2.6$\Mjup.  The rms about the 
two-planet fit is 3.2\,\ms, and there are no significant residual 
periodicities.  The data and two-planet model are shown in 
Figure~\ref{154fit}, and the orbital fits for the individual planets are 
shown in Figure~\ref{154eachone}.

\section{Dynamical Stability Testing}

Recent work has shown that any claim of multiple orbiting bodies must be 
checked by dynamical stability testing to ensure that the proposed 
planetary orbits are feasible on astronomically relevant timescales.  
Such testing can support the orbit fitting results \citep{texas1, 
142paper, NNSer}, place further constraints on the planetary system 
configurations \citep{texas2, subgiants, tan13}, or show that the 
proposed planets cannot exist in or near the nominal best-fit orbits 
\citep{HUAqr, Goz12, HWVir, NSVSpaper, QSVir}.  While HD\,114613 appears 
to be a single-planet system, HD\,154857 hosts two planets which are so 
widely separated (1.3 and 5.4\,AU) that their dynamical interactions 
might be expected to be negligible.  However, for completeness, we 
performed the dynamical analysis as in our previous work 
\citep{142paper}.

We tested the dynamical stability of the HD\,154857 system using the 
Hybrid integrator within the n-body dynamics package \textsc{Mercury} 
\citep{chambers99}.  Holding the initial orbit of the innermost planet 
fixed, we tested 41x41x15x5 grid of ``clones'' spaced evenly across the 
3$\sigma$ range in the outer planet's semi-major axis $a$, eccentricity 
$e$, periastron argument $\omega$, and mean anomaly $M$, respectively.

In each integration, the orbital evolution of each planet was followed 
until it was either ejected from the planetary system (by reaching a 
barycentric distance of 10 AU), or collided with the central body or one 
of the other planets.  The times at which such events occurred was 
recorded, which allowed us to construct a map of the stability of the 
HD\,154857 planetary system as a function of the semi-major axis and 
eccentricity of the outer planet.  As expected, the entire 3$\sigma$ 
region exhibited stability for the full $10^8$yr.  Indeed, not a single 
ejection or collision occurred in any of the 126,075 trial systems.

\section{Discussion and Conclusions}

We have described the detection of two Jupiter-analog planets from the 
15-year Anglo-Australian Planet Search program.  Our new data confirm 
the planetary nature of the previously unconstrained outer body in the 
HD\,154857 system \citep{m04, otoole07}.  These results highlight the 
importance of continuing ``legacy'' programs such as the AAPS, which is 
among the world's longest-running radial-velocity planet searches.  The 
planets detailed in this work bring the total number of AAPS-discovered 
Jupiter analogs to 8.  With three Jupiter analogs confirmed in the past 
two years (HD\,142c, Wittenmyer et al. 2012b; HD\,114613b and 
HD\,154857c, this work), the AAPS has nearly doubled its discoveries of 
these objects in years 14 and 15 of operation.  We expect further 
discoveries of Jupiter analogs over the next few years as additional 
candidates complete orbits.

The AAPS has shifted its primary focus to the search for Jupiter 
analogs.  Central to this strategy is the selection of a subset of 
$\sim$120 targets (from the original 250-star AAPS sample) which satisfy 
two criteria: (1) sufficient observational baseline to detect a Jupiter 
analog, and (2) a sufficiently small velocity scatter to enable the 
robust detection of the $\sim$5-15\,\ms\ signal produced by a Jupiter 
analog.  Criterion (1) eliminates those stars added to the AAPS list 
well after its inception in 1998, and criterion (2) eliminates those 
stars which have high levels of intrinsic activity noise which would 
severely degrade the achievable detection limit.  Using the detections 
and stringent limits from the \textit{non-detections}, for every target 
we will be able to detect or exclude Jupiter analogs with high 
confidence.  The result will be a direct measurement of the frequency of 
such objects, without suffering from significant incompleteness, which 
adds substantial uncertainty to this measurement 
\citep[e.g.][]{cumming08, jupiters}.

There is an emerging correlation between debris disks and low-mass 
planets, first noted by \citet{wyatt12}.  They used \textit{Herschel} to 
detect debris disks around 4 of 6 stars known to host only low-mass 
planets; no debris disks were found in the 5 systems hosting giant 
planets.  One of the stars discussed here, HD\,154857, has been observed 
for infrared excess (indicative of debris disks akin to the Solar 
system's Edgeworth-Kuiper Belt).  No excess was found from 
\textit{Spitzer} and \textit{Herschel} observations 
\citep{bryden09,dod11}.  The HD\,154857 system, hosting two giant 
planets and no detectable debris, is consistent with the pattern noted 
by \citet{wyatt12}.

To obtain a complete picture of the nature of the planet candidates we 
have presented here, it would be ideal to determine true masses, rather 
than the minimum mass derived from radial-velocity measurements.  Direct 
imaging offers a way forward: for stars known to host a long-period 
radial-velocity planet candidate, imaging can determine whether that 
object is stellar (i.e. detectable by imaging) or substellar.  This type 
of characterization has been done for some planet candidates, such as 
14~Herculis~c ($a>7$\,AU -- Wittenmyer et al.~2007), for which AO 
imaging by \citet{rodigas11} established an upper limit of 42\Mjup.  The 
TRENDS survey \citep{crepp12, crepp13a, crepp13b} is currently using 
this strategy to target stars with known radial-velocity trends.  The 
Gemini Planet Imager (GPI), now installed on the 8m Gemini South 
telescope \citep{hartung13}, has been specifically designed for the 
detection of these giant planets \citep{mcbride11}.  It will provide not 
only the high contrasts needed to detect them, but also low-resolution 
spectra for each planet found, which can be used for their 
characterisation.  We are now at a convergence of two developments in 
exoplanetary science: (1) radial-velocity data now extend comfortably 
into the range of Jupiter-analog orbital periods, and (2) direct imaging 
techniques have improved to the point where it is possible to detect 
Jupiter-like planets orbiting Sun-like stars at orbital distances 
approaching that of our own Jupiter ($\sim$5~AU).  These complementary 
techniques can bridge the detectability gap, enabling direct 
measurements of the occurrence rate of Jupiter analogs orbiting Sun-like 
stars.

\acknowledgements

We thank the referee, William Cochran, for a timely report which 
improved this manuscript.  This research is supported by Australian 
Research Council grants DP0774000 and DP130102695.  This research has 
made use of NASA's Astrophysics Data System (ADS), and the SIMBAD 
database, operated at CDS, Strasbourg, France.  This research has also 
made use of the Exoplanet Orbit Database and the Exoplanet Data Explorer 
at exoplanets.org \citep{wright11}.



\begin{deluxetable}{lrr}
\tabletypesize{\scriptsize}
\tablecolumns{3}
\tablewidth{0pt}
\tablecaption{AAT/UCLES Radial Velocities for HD 114613}
\tablehead{
\colhead{JD-2400000} & \colhead{Velocity (\ms)} & \colhead{Uncertainty
(\ms)}}
\startdata
\label{114613vels}
50830.25655  &      4.0  &    1.2  \\
50833.22381  &      1.9  &    0.8  \\
50915.08074  &      0.6  &    1.4  \\
50917.08662  &     -7.5  &    1.6  \\
50970.91681  &    -11.8  &    1.1  \\
51002.86733  &     -5.6  &    1.8  \\
51212.26513  &      5.6  &    1.4  \\
51236.21281  &     -5.5  &    1.4  \\
51237.18130  &     -7.5  &    1.8  \\
51274.21875  &     -1.8  &    2.0  \\
51276.08909  &     -0.2  &    1.3  \\
51382.92476  &     -3.9  &    1.2  \\
51386.85182  &      1.8  &    1.2  \\
51631.24051  &     -1.9  &    1.3  \\
51682.84034  &      1.3  &    1.4  \\
51684.05975  &     -2.8  &    1.4  \\
51717.83623  &     -0.1  &    1.3  \\
51919.25519  &     -0.8  &    1.8  \\
51920.26898  &     -1.7  &    1.5  \\
51984.11127  &      0.3  &    1.7  \\
52061.03012  &      5.0  &    1.4  \\
52062.09081  &      3.8  &    1.5  \\
52092.95741  &      3.6  &    1.1  \\
52127.88681  &      0.4  &    1.7  \\
52359.20083  &      2.3  &    1.0  \\
52387.02810  &      4.4  &    1.4  \\
52388.06604  &      3.4  &    1.5  \\
52509.86723  &     13.4  &    1.3  \\
52510.86704  &     11.2  &    1.5  \\
52654.26882  &      4.6  &    1.4  \\
52710.14699  &      8.8  &    1.0  \\
52710.95867  &      6.4  &    2.0  \\
52745.09064  &     16.9  &    1.3  \\
52751.11506  &      9.0  &    1.4  \\
52752.07289  &     11.6  &    1.1  \\
52783.95975  &     11.8  &    1.2  \\
52785.05762  &      9.3  &    1.7  \\
52785.97105  &     15.2  &    1.4  \\
52857.87314  &      0.6  &    1.3  \\
53008.22116  &     11.7  &    1.3  \\
53041.22754  &      8.0  &    1.3  \\
53042.22343  &      4.2  &    1.2  \\
53046.15053  &      5.2  &    1.5  \\
53051.15875  &     -0.3  &    1.0  \\
53214.87828  &     10.8  &    1.0  \\
53215.88215  &      9.6  &    1.2  \\
53242.89915  &      4.5  &    0.8  \\
53245.84896  &     14.4  &    1.5  \\
53399.27632  &      5.5  &    0.6  \\
53405.20308  &      2.0  &    0.7  \\
53482.95081  &     -1.3  &    0.8  \\
53484.04090  &     -4.9  &    0.7  \\
53485.02218  &     -0.9  &    0.7  \\
53485.94007  &      0.0  &    0.7  \\
53486.06578  &     -0.8  &    0.7  \\
53488.12718  &      1.2  &    0.7  \\
53489.07405  &     -0.4  &    0.7  \\
53506.95930  &      1.5  &    0.7  \\
53507.88240  &      0.4  &    0.7  \\
53509.07069  &     -4.2  &    0.7  \\
53516.02470  &      2.0  &    1.0  \\
53517.00282  &      6.4  &    0.8  \\
53518.95553  &      6.2  &    0.7  \\
53520.02670  &      6.0  &    0.8  \\
53521.01119  &      7.7  &    0.8  \\
53521.93055  &      4.4  &    0.9  \\
53522.97922  &      4.0  &    0.8  \\
53568.93024  &      1.4  &    0.7  \\
53569.89040  &     -2.5  &    0.8  \\
53570.93229  &     -0.1  &    0.7  \\
53571.92818  &      4.4  &    0.7  \\
53572.93960  &      3.5  &    0.6  \\
53573.86024  &      2.8  &    0.7  \\
53575.86501  &      4.8  &    0.6  \\
53576.83855  &      1.5  &    0.6  \\
53577.85712  &      0.4  &    0.9  \\
53578.87161  &      1.6  &    0.6  \\
53840.18077  &     -2.4  &    0.9  \\
53841.14463  &      2.1  &    0.8  \\
53843.11871  &      0.4  &    0.8  \\
53844.06189  &      1.7  &    0.7  \\
53937.92281  &     -1.6  &    0.7  \\
53938.90015  &      1.4  &    0.6  \\
53943.85615  &     -4.9  &    0.6  \\
53944.87540  &     -7.3  &    0.6  \\
53945.86861  &     -4.3  &    0.8  \\
53946.86415  &     -7.3  &    0.6  \\
54111.19771  &      3.5  &    0.6  \\
54112.20754  &      4.7  &    0.8  \\
54113.22282  &      3.5  &    0.8  \\
54114.24651  &      2.6  &    0.9  \\
54115.25260  &      3.5  &    1.2  \\
54120.20490  &      7.7  &    0.7  \\
54121.19705  &      3.8  &    0.6  \\
54123.22186  &      9.1  &    0.6  \\
54126.18385  &      2.7  &    0.7  \\
54127.18893  &      5.0  &    0.5  \\
54128.18767  &      4.0  &    0.8  \\
54129.18497  &     -0.7  &    0.5  \\
54130.17738  &      0.9  &    0.7  \\
54131.18561  &      2.0  &    0.6  \\
54132.19227  &      2.1  &    0.8  \\
54133.25166  &      1.4  &    1.1  \\
54134.21614  &      4.1  &    0.6  \\
54135.18151  &     -1.1  &    0.7  \\
54136.20085  &      0.8  &    0.6  \\
54137.19842  &      1.3  &    0.6  \\
54138.17883  &      0.1  &    0.9  \\
54139.17104  &      3.0  &    0.8  \\
54140.17481  &     -0.1  &    0.8  \\
54141.20213  &     -1.6  &    0.8  \\
54142.19235  &      0.6  &    0.5  \\
54144.12894  &     -3.4  &    0.6  \\
54145.15798  &     -2.9  &    0.6  \\
54146.17863  &     -0.3  &    0.6  \\
54147.19558  &     -3.5  &    0.6  \\
54148.22623  &     -2.7  &    0.6  \\
54149.16439  &     -2.0  &    0.6  \\
54150.19348  &     -1.7  &    0.6  \\
54151.20779  &     -3.0  &    0.6  \\
54152.22033  &     -2.7  &    0.7  \\
54154.19334  &     -5.2  &    0.6  \\
54155.19519  &     -3.2  &    0.7  \\
54156.16473  &     -1.3  &    0.6  \\
54223.10365  &      1.9  &    0.9  \\
54224.14365  &      1.1  &    0.8  \\
54225.07630  &      1.8  &    0.7  \\
54226.00805  &      1.9  &    1.0  \\
54227.04258  &      5.5  &    0.8  \\
54252.96645  &      9.5  &    0.9  \\
54254.00991  &      9.0  &    0.8  \\
54254.91285  &      5.6  &    0.9  \\
54255.97533  &      3.7  &    0.9  \\
54257.06303  &      5.6  &    0.9  \\
54333.85357  &      3.5  &    0.9  \\
54334.87415  &      2.6  &    0.9  \\
54335.86542  &      2.5  &    0.7  \\
54336.85846  &      6.7  &    0.8  \\
54543.05240  &      0.0  &    0.8  \\
54544.14396  &     -0.8  &    0.8  \\
54550.09556  &     -1.0  &    1.4  \\
54551.09203  &     -1.9  &    0.9  \\
54552.13245  &      3.0  &    1.1  \\
54553.10244  &     -3.6  &    1.1  \\
54841.25120  &     -0.4  &    1.0  \\
54897.20442  &     -8.4  &    1.0  \\
54900.18137  &    -12.3  &    1.2  \\
54901.15823  &     -6.8  &    0.9  \\
54902.20552  &    -10.1  &    1.0  \\
54904.19780  &     -6.0  &    1.0  \\
54905.26523  &     -1.6  &    0.8  \\
54906.21006  &     -5.8  &    1.0  \\
54907.19989  &     -6.8  &    0.9  \\
54908.20498  &     -6.8  &    0.8  \\
55014.94187  &     -6.2  &    1.0  \\
55015.86595  &     -4.9  &    0.8  \\
55018.84451  &      1.4  &    1.9  \\
55019.87410  &     -5.8  &    0.7  \\
55020.84762  &     -6.3  &    0.7  \\
55021.87399  &     -7.7  &    0.8  \\
55022.88779  &    -10.4  &    0.8  \\
55023.86739  &    -11.0  &    0.9  \\
55029.84869  &     -6.6  &    0.6  \\
55030.83778  &     -4.0  &    0.7  \\
55031.90051  &     -3.4  &    0.7  \\
55032.91264  &     -3.6  &    0.6  \\
55036.84785  &     -6.2  &    0.7  \\
55037.83638  &     -2.9  &    0.7  \\
55040.84113  &     -6.1  &    0.8  \\
55041.85626  &     -6.0  &    0.9  \\
55043.88548  &     -3.7  &    0.6  \\
55044.85993  &     -4.9  &    0.6  \\
55045.85496  &     -5.0  &    0.5  \\
55046.90149  &     -6.5  &    0.7  \\
55047.88023  &     -4.0  &    0.5  \\
55048.86964  &     -5.4  &    0.5  \\
55049.85230  &     -4.1  &    1.0  \\
55050.85220  &     -7.8  &    0.8  \\
55051.84966  &     -5.0  &    0.5  \\
55053.85289  &     -8.7  &    0.6  \\
55054.84736  &     -8.0  &    0.6  \\
55055.88706  &     -6.2  &    0.5  \\
55058.87417  &     -4.9  &    0.6  \\
55206.19055  &     -2.4  &    0.8  \\
55209.20056  &     -1.9  &    0.8  \\
55253.19939  &     -6.9  &    0.9  \\
55254.28349  &     -9.6  &    1.0  \\
55310.06142  &      2.1  &    0.8  \\
55312.07035  &     -2.4  &    0.8  \\
55317.03644  &     -5.0  &    0.9  \\
55370.94153  &      3.1  &    1.0  \\
55374.90773  &      0.7  &    1.0  \\
55397.89284  &     -6.3  &    1.0  \\
55402.84315  &     -5.0  &    0.7  \\
55586.21266  &      0.8  &    0.9  \\
55603.27577  &      3.8  &    1.1  \\
55604.27940  &     -0.2  &    1.1  \\
55663.95088  &      0.5  &    1.1  \\
55666.06790  &      2.5  &    0.9  \\
55692.08906  &      0.5  &    1.1  \\
55692.97915  &      5.8  &    0.9  \\
55750.89289  &     -3.6  &    0.8  \\
55751.87344  &     -2.2  &    1.3  \\
55753.84964  &     -3.0  &    1.0  \\
55756.87098  &     -1.2  &    1.1  \\
55785.90110  &      2.3  &    1.0  \\
55786.86812  &      4.9  &    1.5  \\
55787.92128  &      5.1  &    1.1  \\
55961.17818  &     -0.1  &    0.8  \\
55964.22153  &      6.2  &    0.8  \\
55996.09776  &     -3.3  &    0.9  \\
56049.03768  &      2.1  &    1.1  \\
56051.03503  &     -0.1  &    1.1  \\
56084.99567  &      9.3  &    1.3  \\
56085.98717  &      3.6  &    1.1  \\
56134.98050  &      6.9  &    1.2  \\
56138.89373  &      3.4  &    1.1  \\
56379.13086  &      2.6  &    1.3  \\
56382.21632  &     10.6  &    1.9  \\
56383.07275  &      1.5  &    0.8  \\
56464.91329  &     10.2  &    1.1  \\
56465.92701  &      9.3  &    0.9  \\
56466.91347  &      8.0  &    0.9  \\
\enddata
\end{deluxetable}


\begin{deluxetable}{lrr}
\tabletypesize{\scriptsize}
\tablecolumns{3}
\tablewidth{0pt}
\tablecaption{AAT/UCLES Radial Velocities for HD 154857}
\tablehead{
\colhead{JD-2400000} & \colhead{Velocity (\ms)} & \colhead{Uncertainty
(\ms)}}
\startdata
\label{154857vels}
52389.23580  &     -3.3  &    1.7  \\
52390.21223  &     -4.9  &    1.5  \\
52422.13713  &    -17.3  &    1.5  \\
52453.01957  &    -14.2  &    1.4  \\
52455.02535  &    -13.7  &    1.9  \\
52455.97664  &    -12.2  &    1.9  \\
52509.94853  &    -16.3  &    2.1  \\
52510.91619  &     -6.6  &    1.8  \\
52711.24602  &     63.7  &    2.8  \\
52745.24271  &     66.4  &    1.9  \\
52747.21175  &     58.7  &    1.8  \\
52750.17761  &     52.7  &    1.5  \\
52751.22944  &     47.2  &    1.4  \\
52784.12626  &    -11.5  &    1.3  \\
52857.02974  &    -28.4  &    2.7  \\
52857.98599  &    -31.9  &    1.4  \\
52942.91225  &    -18.1  &    1.7  \\
53217.01252  &    -41.2  &    1.5  \\
53246.03809  &    -53.3  &    1.9  \\
53485.15229  &     21.5  &    1.5  \\
53510.15968  &     25.9  &    1.4  \\
53523.10133  &     33.2  &    1.5  \\
53570.02945  &     32.9  &    2.3  \\
53843.23961  &     -7.0  &    1.6  \\
53945.03237  &     49.6  &    1.0  \\
54008.89626  &    -21.4  &    0.9  \\
54037.88134  &    -39.5  &    1.4  \\
54156.28808  &    -17.5  &    1.9  \\
54226.21678  &     -2.3  &    1.2  \\
54254.99229  &      5.5  &    1.6  \\
54372.93185  &     61.1  &    1.3  \\
54552.23398  &     -9.0  &    2.2  \\
54901.28209  &     -7.9  &    2.1  \\
55101.89348  &     54.8  &    1.7  \\
55313.23386  &      2.5  &    1.4  \\
55317.13834  &      1.1  &    1.7  \\
55399.04324  &     17.8  &    1.4  \\
55429.91942  &     23.9  &    1.3  \\
56049.24780  &      1.1  &    2.1  \\
56465.07924  &    -22.4  &    1.9  \\
56467.06137  &    -20.2  &    1.7  \\
56498.02816  &    -40.6  &    3.0  \\
\enddata
\end{deluxetable}

\begin{deluxetable}{lll}
\tabletypesize{\scriptsize}
\tablecolumns{3}
\tablewidth{0pt}
\tablecaption{Stellar Parameters for HD 114613 }
\tablehead{
\colhead{Parameter} & \colhead{Value} & \colhead{Reference}
 }
\startdata
\label{114star}
Spec.~Type & G4IV & \citet{gray06} \\
  & G3V & \citet{torres06} \\
Mass (\Msun) & 1.364 & \citet{sousa08} \\ 
Distance (pc) & 20.67$\pm$0.13 & \citet{vl07} \\
V sin $i$ (\kms) & 2.7$\pm$0.9 & \citet{saar97} \\ 
log$R'_{HK}$ & -5.118 & \citet{gray06} \\
$[Fe/H]$ & 0.19$\pm$0.01 & \citet{sousa08} \\
  & 0.18 & \citet{mal12} \\
$[O/H]$ & 0.03$\pm$0.01 & \citet{bond08} \\
$[Cr/H]$ & 0.09$\pm$0.04 & \citet{bond08} \\
$[Mg/H]$ & -0.04$\pm$0.06 & \citet{bond08} \\
$[Zr/H]$ & 0.17$\pm$0.04 & \citet{bond08} \\
$[Eu/H]$ & -0.17$\pm$0.04 & \citet{bond08} \\
$[Nd/H]$ & 0.05$\pm$0.01 & \citet{bond08} \\
$[Si/H]$ & 0.19$\pm$0.06 & \citet{bond06} \\  
$T_{eff}$ (K) & 5729$\pm$17 & \citet{sousa08} \\
  & 5574 & \citet{gray06} \\
  & 5550 & \citet{saar97} \\
log $g$ & 3.97$\pm$0.02 & \citet{sousa08} \\
  & 3.90 & \citet{gray06} \\
\enddata
\end{deluxetable}


\begin{deluxetable}{lll}
\tabletypesize{\scriptsize}
\tablecolumns{3}
\tablewidth{0pt}
\tablecaption{Stellar Parameters for HD 154857 }
\tablehead{
\colhead{Parameter} & \colhead{Value} & \colhead{Reference}
 }
\startdata
\label{154star}
Spec.~Type & G5V & \citet{houk75} \\
Mass (\Msun) & 1.718$^{+0.03}_{-0.022}$ & \citet{takeda07} \\ 
Distance (pc) & 64.2$\pm$3.1 & \citet{vl07} \\
V sin $i$ (\kms) & 1.4$\pm$0.5 & \citet{butler06} \\ 
log$R'_{HK}$ & -5.00 & \citet{jenkins06} \\
  & -5.14 & \citet{henry96} \\
$[Fe/H]$ & -0.31 & \citet{h09} \\
  & -0.30 & \citet{g10} \\
  & -0.22 & \citet{vf05} \\
  & -0.20 & \citet{casa11} \\
$[O/H]$ & -0.15$\pm$0.03 & \citet{bond08} \\
$[Cr/H]$ & -0.20$\pm$0.04 & \citet{bond08} \\
$[Mg/H]$ & -0.20$\pm$0.03 & \citet{bond08} \\
$[Zr/H]$ & -0.08$\pm$0.04 & \citet{bond08} \\
$[Eu/H]$ & -0.27$\pm$0.07 & \citet{bond08} \\
$[Nd/H]$ & -0.01$\pm$0.02 & \citet{bond08} \\
$[C/H]$ & -0.28$\pm$0.07 & \citet{bond06} \\
$[Si/H]$ & -0.28$\pm$0.11 & \citet{bond06} \\
$T_{eff}$ (K) & 5605 & \citet{dod11} \\
  & 5508 & \citet{h09} \\
  & 5548 & \citet{g10} \\
log $g$ & 3.95$^{+0.05}_{-0.03}$ & \citet{takeda07} \\
  & 3.82 & \citet{g10} \\
  & 3.99$\pm$0.06 & \citet{vf05} \\
Radius (\Rsun) & 2.31$^{+0.17}_{-0.10}$ & \citet{takeda07} \\
  & 1.760$\pm$0.057 & \citet{torres10} \\
\enddata
\end{deluxetable}

\begin{deluxetable}{lr@{$\pm$}lr@{$\pm$}lr@{$\pm$}lr@{$\pm$}lr@{$\pm$}lr@{$\pm$}
lr@{$\pm$}ll}
\tabletypesize{\scriptsize}
\tablecolumns{11}
\tablewidth{0pt}
\tablecaption{Keplerian Orbital Solutions }
\tablehead{
\colhead{Planet} & \multicolumn{2}{c}{Period} & \multicolumn{2}{c}{$T_0$}
&
\multicolumn{2}{c}{$e$} & \multicolumn{2}{c}{$\omega$} &
\multicolumn{2}{c}{K } & \multicolumn{2}{c}{m sin $i$ } &
\multicolumn{2}{c}{$a$ } & \colhead{$\chi^2_{\nu}$} \\
\colhead{} & \multicolumn{2}{c}{(days)} & \multicolumn{2}{c}{(JD-2400000)}
&
\multicolumn{2}{c}{} &\multicolumn{2}{c}{(degrees)} & \multicolumn{2}{c}{(\ms)} &
\multicolumn{2}{c}{(\Mjup)} & \multicolumn{2}{c}{(AU)} & 
 }
\startdata
\label{planetparams}   
HD 114613 b & 3827 & 105 & 55550.3 & (fixed) & 0.25 & 0.08 & 244 & 5 &
5.52 & 0.40 & 0.48 & 0.04 & 5.16 & 0.13 & 1.23 \\
HD 154857 b & 408.6 & 0.5 & 53572.5 & 2.4 & 0.46 & 0.02 & 57 & 4 &
48.3 & 1.0 & 2.24 & 0.05 & 1.291 & 0.008 & 1.35 \\
HD 154857 c & 3452 & 105 & 55219 & 375 & 0.06 & 0.05 & 352 & 37 &
24.2 & 1.1 & 2.58 & 0.16 & 5.36 & 0.09 & \\
\enddata
\end{deluxetable}

\begin{deluxetable}{llll}
\tabletypesize{\scriptsize}
\tablecolumns{4}
\tablewidth{0pt}
\tablecaption{Candidate Secondary Signals for HD\,114613 }
\tablehead{
\colhead{First Half} & \colhead{} & \colhead{Second Half} & \colhead{} \\
\colhead{Period (days)} & \colhead{FAP} & \colhead{Period (days)} &
\colhead{FAP}
 }
\startdata
\label{boot}
28.9 & 0.015 & 26.6 & 0.0001 \\
72.5 & 0.952 & 73.3 & 0.1836 \\
122 & 1.000 & 121.4 & 0.0246 \\
490.2 & 0.138 & 480.8 & 0.9755 \\
1562.5 & 0.910 & 1111.1 & 0.0003 \\
\enddata
\end{deluxetable}

\clearpage

\end{document}